# Model-assisted cohort selection with bias analysis for generating large-scale cohorts from the EHR for oncology research

Benjamin Birnbaum*[1], Nathan Nussbaum*[1,2], Katharina Seidl-Rathkopf*[1], Monica Agrawal[13], Melissa Estévez[1], Evan Estola[1], Joshua Haimson[1], Lucy He[1], Peter Larson[1], Paul Richardson[1]

*Equal contribution

[1]Flatiron Health Inc.
New York, NY, USA

[2]New York University School of Medicine
New York, NY, USA

[3]Massachusetts Institute of Technology
Cambridge, MA, USA

**Corresponding Author:**
Benjamin Birnbaum
233 Spring St., Fifth Floor
New York, NY 10013
USA
bbirnbaum@flatiron.com
503-702-8123

# Abstract

**OBJECTIVE**

Electronic health records (EHRs) are a promising source of data for health outcomes research in oncology. A challenge in using EHR data is that selecting cohorts of patients often requires information in unstructured parts of the record. Machine learning has been used to address this, but even high-performing algorithms may select patients in a non-random manner and bias the resulting cohort. To improve the efficiency of cohort selection while measuring potential bias, we introduce a technique called Model-Assisted Cohort Selection (MACS) with Bias Analysis and apply it to the selection of metastatic breast cancer (mBC) patients.

**MATERIALS AND METHODS**

We trained a model on 17,263 patients using term-frequency inverse-document-frequency (TF-IDF) and logistic regression. We used a test set of 17,292 patients to measure algorithm performance and perform Bias Analysis. We compared the cohort generated by MACS to the cohort that would have been generated without MACS as reference standard, first by comparing distributions of an extensive set of clinical and demographic variables and then by comparing the results of two analyses addressing existing example research questions.

**RESULTS**

Our algorithm had an area under the curve (AUC) of 0.976, a sensitivity of 96.0%, and an abstraction efficiency gain of 77.9%. During Bias Analysis, we found no large differences in baseline characteristics and no differences in the example analyses.

**CONCLUSION**

MACS with bias analysis can significantly improve the efficiency of cohort selection on EHR data while instilling confidence that outcomes research performed on the resulting cohort will not be biased.

**INTRODUCTION**

The widespread adoption of electronic health records (EHRs) by oncology practices in recent years has enabled EHRs to become an important source of data for research. Like other types of real-world data (RWD), EHR data are generated from routine clinical care and describe broad patient populations. In oncology specifically, where the treatment landscape is evolving rapidly, the use of deidentified RWD for research provides an opportunity to advance knowledge faster than allowed by prospective clinical trials alone. Compared to other RWD sources like registries and claims datasets, EHRs offer advantages for research such as longitudinality of data collection, recency, generalizability, and depth of clinical information capture.[1]

Repurposing EHR data for research presents unique challenges but also introduces opportunities to use technology to accelerate research. One fundamental challenge is that important information is often not organized into a consistent data model but is instead distributed across unstructured sources like clinical notes and pathology reports.[1] Due to the complexity of oncology records, this process of transitioning unstructured data into structured variables currently requires manual chart review by trained human abstractors. Chart review, however, can be slow, costly, and lack quality control.[2] To address these obstacles, we have developed a technique called technology-enabled abstraction, aimed at increasing the accuracy and efficiency with which human abstractors can extract critical data elements from each chart.[1, 3] Even with this technique, abstraction remains time intensive.

One challenge closely related to unstructured data is the selection of research cohorts from EHR data, which must be performed according to predefined inclusion and exclusion criteria. Many of these criteria, ranging from disease features like histology type to patient characteristics like genetic predisposition, are not well-captured in structured data alone. As a result, relying solely on structured data for cohort selection can miss patients.[1] While technology-enabled abstraction can make the use of unstructured data for cohort selection more efficient, the process does not eliminate the need to review all potentially cohort-eligible patients when creating a new research dataset. Therefore, better approaches to performing cohort selection at scale are needed.

Previous work has explored using machine learning (ML) and rule-based systems, often in combination with natural language processing (NLP), for creating cohorts based on EHR data in an attempt to decrease the burden of manual chart review (see e.g., [4-6]; for a review, see [7-10]). While high-performing algorithms are promising, they introduce the risk of selecting patients in a non-random manner. This could cause the resulting cohort to be biased, which could lead to erroneous research findings.

Here we introduce a framework called Model-Assisted Cohort Selection (MACS) with Bias Analysis to more efficiently identify cohort-eligible patients and to assess the resulting cohort's suitability for research. With MACS, potentially eligible patients pass through a two-step cohort-selection filter (Figure 1). In the first step, a model predicts which patients are eligible for cohort inclusion. In the second step, human abstractors manually assess patients deemed cohort-eligible by the model for cohort eligibility. Then Bias Analysis compares the cohort produced by MACS to a reference cohort in which every inclusion or exclusion decision was made by a human abstractor. In addition to standard measures such as sensitivity and specificity, this comparison includes the distributions of an extensive set of clinical and demographic variables as well as the results of analyses addressing example research questions.

This paper illustrates MACS with Bias Analysis by applying it to the selection a cohort of metastatic breast cancer (mBC) patients. Even though ICD codes exist that could be used to identify patients with metastatic disease, these codes have been shown to lack sensitivity compared to other methods.[3, 11] Thus, accurately knowing whether a patient has developed metastatic disease requires using unstructured data,[1] presenting an opportunity for significant efficiency improvement via automation.

## Materials and methods

### Data Source and Processing

The data source for this study was the Flatiron Health database, a longitudinal, demographically and geographically diverse database derived from EHR data. The cutoff date for this study was February 1, 2018. At the time of this study, the database included over 265 cancer clinics (at approximately 800 sites of care) in the United States representing more than two million active patients available for analysis, the majority of whom are seen in the community oncology setting. Institutional Review Board approval was obtained prior to study conduct, and informed patient consent was waived.

Patient-level demographic, clinical, and outcomes data were extracted from the source EHR system and include structured data as well as data elements extracted from unstructured data in clinical documents. Structured data were aggregated, normalized, and harmonized across clinics. Documents were classified into a common set of 24 standard categories. Unstructured data elements were then extracted using technology-enabled abstraction.[1, 3] Dates of death were obtained from a composite mortality variable.[12] Lines of therapy were constructed based on drug order and administration data following oncologist-defined rules.

### Labeled Data Preparation

To generate labeled data for model training, evaluation, and Bias Analysis, we first identified patients with recent activity in the EHR by selecting patients with at least two recorded office visits on separate dates after January 1, 2011. We then identified patients with a likely breast cancer diagnosis based on the presence of a related ICD code (ICD-9 174.x or 175.x or ICD-10 C50.x) in their structured records. We sent a random subset of these patients for chart review by oncology nurses and tumor registrars, who used our in-house technology-enabled abstraction platform [1, 3] to extract a number of variables from the unstructured EHR data, including whether each patient had mBC (defined as Stage IV breast cancer or recurrence of early-stage breast cancer). We then removed patients whose date of metastatic diagnosis was prior to January 1, 2011. We did this to avoid introducing immortal time bias caused by the requirement of having a visit after that date.

This process resulted in a set of 34,555 patients. We randomly selected 50% of these patients for model training (N = 17,263) and reserved the remainder (N = 17,292) as a test set for assessing model performance and evaluating the presence of bias caused by using MACS.

**Modeling approach**

Labels

We assigned a binary label to each patient in the training and test sets: *positive* patients were those patients who had a diagnosis of mBC, and *negative* patients were those who did not have a diagnosis of mBC. Because all patients in this dataset had an ICD code indicating a diagnosis of breast cancer, most of the negative patients had (non-metastatic) breast cancer. Positive patients accounted for approximately 8.5%–9% of cases in both the training and test sets (Table 1).

*Table 1: Number of positive and negative patients in the training and test sets*

|  | **Positive patients** (patients with a chart-confirmed mBC diagnosis after 1/1/2011) | **Negative patients** (patients without evidence of mBC diagnosis based on chart review) |
|---|---|---|
| **Test set** | 1485 | 15807 |
| **Training set** | 1473 | 15790 |

Feature generation

The input to our model was a vector of features derived from unstructured EHR data. We first selected all documents matching four relevant categories: visit notes, pathology reports, procedure and operative reports, and radiology reports. Second, for each patient, we sorted these documents chronologically and concatenated them into a single block of text. Third, we removed special characters and tokenized the text on white space. Fourth, we extracted unigrams and bigrams from this normalized text and weighted them using term-frequency inverse-document-frequency (TF-IDF). Finally, we retained only the 100,000 most commonly occurring features for model training efficiency. In our experiments using cross-validation on the training data, performance did not increase noticeably when including more features.

Training

We randomly selected 80% of our training data and used the labels and features described above to train a logistic regression model. This model used L2-regularization, where the regularization constant was selected to maximize the area under the receiver operating characteristic (ROC) curve during cross-validation on this subset of the training data. We performed feature generation and training using the scikit-learn library.[13]

Choosing a score threshold

The trained logistic regression model outputs a score between 0 and 1 for each patient; a higher score indicates increased confidence that the patient is metastatic. In order to turn this into a binary classifier, a threshold must be chosen such that patients will be classified as metastatic if and only if the model outputs a score higher than this threshold. Using a validation set consisting of the remaining 20% of the training data, we selected a score threshold predicted to give 95% sensitivity. More precisely, we chose the largest value $x$ such that at least 95% of the positive examples in the validation set had a score higher than $x$. We chose this high value for the target sensitivity to reduce the potential for introducing bias.

**Analysis**

Model performance

We assessed performance of the model on the test set by plotting the ROC curve, measuring the area under this curve (AUC), as well as the sensitivity and abstraction efficiency gain of the model with the chosen threshold. We defined abstraction efficiency gain to be the fraction of patients excluded from manual abstraction in the second step of MACS because their scores fell below the threshold. Although we chose a threshold to achieve a certain sensitivity, it is still important to measure sensitivity on the test set to ensure that the chosen threshold generalizes to new data. Confidence intervals for sensitivity and abstraction efficiency were generated based on 1000 bootstrapped samples.

Bias Analysis

To assess whether using MACS introduced bias, we used the set of all positive patients in the test set as a Reference Standard. Because every patient in the test set was manually abstracted to determine metastatic status, the Reference Standard represents the cohort that would have been generated without using MACS. We compared the Reference Standard to the subset of these patients classified as positive by the model. This latter set of patients, called the MACS Cohort, represents the cohort that would have been generated by the MACS process. We used these cohorts to assess MACS for bias using the following two approaches.

First, we compared baseline demographic and clinical distributions in the MACS Cohort to the corresponding distributions in the Reference Standard. We summarized a number of demographic and clinical characteristics within the two cohorts using descriptive statistics (medians and interquartile ranges (IQR) for continuous variables and percentages for categorical variables). For each measure of interest, we calculated the difference between the two cohorts and obtained 95% empirical confidence intervals (CIs) using the percentile method by drawing 1000 bootstrap samples with replacement equal in size to the two original cohorts.

Our second approach compared the results of analyses addressing existing example research questions on the MACS Cohort to the corresponding results of analyses on the Reference Standard. The rationale for this approach was that if the MACS Cohort is unbiased, results from any analysis conducted in the Reference Standard should be replicable in the MACS Cohort. The example analyses were chosen to be representative of the types of analyses that will ultimately be conducted on the cohort. In this work, we illustrate the idea with two basic outcomes analyses: (1) What is the median overall survival (OS) following diagnosis with mBC, and (2) how does OS vary by hormone receptor (HR) and human epidermal growth factor receptor 2 (HER2) status at the time of mBC diagnosis? HR status is defined as negative if the tumor is found to express neither progesterone receptors (PRs) nor estrogen receptors (ERs) and defined as positive for any other finding. HR and HER2 status are important prognostic factors in mBC and inform treatment decisions. We specifically compared OS in patients with negative HR and negative HER2 status (triple negative) to OS in patients with positive HR status and negative HER2 status (HR+/HER2-) as triple negative breast cancer is known to have worse outcomes than HR+/HER2- breast cancer. [14] We extracted HR and HER2 status at the time of metastatic diagnosis via technology-enabled chart abstraction of relevant information.[1]

We performed survival analyses using standard Kaplan-Meier methods [15]. We defined the index date to be the date of a patient's metastatic diagnosis and the event date to be the date of death due to any cause. We censored patients who were lost to follow-up or had not yet experienced the relevant event by the end of the study period at the date of last confirmed activity (last clinical visit or drug administration date). We calculated medians, 95% CIs, and Kaplan-Meier curves for each outcome. For each comparison we further calculated the difference in median OS between the MACS Cohort and the Reference Standard, computing 95% empirical CIs for the difference with the bootstrapping method described above. In the analysis of OS by HR and HER2 status we further compared OS between mBC subpopulations (HR+/HER2- vs. triple negative) within the same cohort (Reference Standard or MACS Cohort) for statistically significant differences using log rank tests.

---

[1] We identified ER, PR, and HER2 status at time of metastatic diagnosis as follows: the closest successful test within 60 days of metastatic diagnosis date or when no such test is available, the closest successful test before metastatic diagnosis date. When only tests with pending, inconclusive, or unknown results were available, we applied the same date criteria to select the most relevant test. Patients were considered to be HR+ if they had either a positive ER or a positive PR status based on the above definition

## Results

**Model performance**

Our logistic regression model for predicting whether a patient was diagnosed with mBC had an AUC on our held out test data of 0.976 (95% CI: 0.973, 0.979). The ROC curve of the model is shown in Figure 2. The score threshold chosen to achieve 95% sensitivity during model training was 0.047. Sensitivity in the test set was 96.0% (95% CI: 94.9, 96.9), indicating that the threshold chosen during model training generalized well. The abstraction efficiency gain for the chosen threshold was 77.9% (95% CI: 77.2, 78.5). This means that the model reduced the number of charts needing human review by 78% while missing 4% of the positive patients (Table 2).

*Table 2: comparison of cohort selection using a standard technology-enabled abstraction approach vs MACS.*

|  | Standard approach | MACS | Difference | Decrease due to MACS |
|---|---|---|---|---|
| **Patients abstracted** | 17292 | 3827 | 13465 | 77.87% |
| **Patients in resulting cohort** | 1485 | 1425 | 60 | 4.04% |

**Bias Analysis**

Distribution of demographic and clinical characteristics

We compared distributions of demographic and clinical baseline characteristics to assess if missing 4% of patients through MACS results in meaningful changes that could bias the results of downstream analyses. Table 3 shows the comparison of demographic and clinical characteristics between the Reference Standard and the MACS Cohort. Overall, differences between the two cohorts were small; for example, median age at initial and metastatic diagnosis and median follow-up time differed by less than a year between the two cohorts. For all measured categorical variables and levels, the observed differences between the two cohorts were less than 1% and the upper and lower bounds suggested by the 95% CIs were all less than 4%. Qualitatively, the biggest differences were observed in the year of metastatic diagnosis date, with the MACS cohort showing a small shift towards earlier diagnosis dates, and in the stage at primary breast cancer diagnosis, with the MACS cohort showing a shift toward Stage IV.

*Table 3: Comparison of demographic and clinical baseline characteristics. The rightmost column shows the difference between the Reference Standard and the MACS Cohort (in percent for categorical variables and medians for continuous variables).*

|  | Ref. Standard (N = 1485) | MACS Cohort (N = 1425) | Ref. Standard - MACS Cohort [CI] |
|---|---|---|---|
| **Demographics** | | | |
| Median age at primary breast cancer diagnosis, years [IQR] | 59 [50, 68] | 59 [50, 68] | 0 [-2, 1] |
| Age at primary breast cancer diagnosis category, n [%] | | | |
|   19-34 | 47 [3.16] | 47 [3.3] | -0.13 [-1.04, 1.36] |
|   35-50 | 317 [21.35] | 308 [21.61] | -0.27 [-2.74, 3.33] |
|   50-64 | 595 [40.07] | 575 [40.35] | -0.28 [-3.43, 3.79] |
|   65-74 | 313 [21.08] | 292 [20.49] | 0.59 [-3.53, 2.49] |
|   75+ | 200 [13.47] | 191 [13.40] | 0.06 [-2.63, 2.23] |
|   Unknown | 13 [0.88] | 12 [0.84] | 0.03 [-0.72, 0.59] |
| Median age at mBC diagnosis, years [IQR] | 64 [55, 73] | 64 [55, 73] | 0 [-2, 1] |
| Age at mBC diagnosis category, n [%] | | | |
| 19-34 | 24 [1.62] | 24 [1.68] | -0.07 [-0.82, 1.03] |
| 35-50 | 190 [12.79] | 184 [12.91] | -0.12 [-2.23, 2.45] |
| 50-64 | 541 [36.43] | 524 [36.77] | -0.34 [-2.99, 3.62] |
| 65-74 | 394 [26.53] | 375 [26.32] | 0.22 [-3.41, 2.85] |
| 75+ | 336 [22.63] | 318 [22.32] | 0.31 [-3.33, 2.58] |
| Gender, n [%] | | | |
| Female | 1474 [99.26] | 1415 [99.30] | -0.04 [-0.52, 0.65] |
| Male | 11 [0.74] | 10 [0.70] | 0.04 [-0.65, 0.52] |
| Race/ethnicity, n [%] | | | |
| White | 958 [64.51] | 915 [64.21] | 0.3 [-3.64, 3.42] |
| Black or African American | 171 [11.52] | 164 [11.51] | 0.01 [-2.39, 2.34] |
| Asian | 46 [3.10] | 46 [3.23] | -0.13 [-1.11, 1.37] |
| Other race | 169 [11.38] | 166 [11.65] | -0.27 [-2.09, 2.46] |
| Unknown | 141 [9.49] | 134 [9.40] | 0.09 [-2.37, 1.84] |
| Practice type, n [%] | | | |
| Academic | 99 [6.67] | 89 [6.25] | 0.42 [-2.06, 1.35] |
| Community | 1386 [93.33] | 1336 [93.75] | -0.42 [-1.35, 2.06] |
| **Clinical characteristics** | | | |
| Stage at primary breast cancer diagnosis, n [%] | | | |
| 0 | < 5 | < 5 | 0 [-0.2, 0.21] |
| I | 136 [9.16] | 127 [8.91] | 0.25 [-2.07, 1.83] |
| II | 356 [23.97] | 334 [23.44] | 0.53 [-3.52, 2.72] |
| III | 311 [20.94] | 297 [20.84] | 0.1 [-2.94, 2.98] |
| IV | 461 [31.04] | 452 [31.72] | -0.68 [-2.66, 4.14] |
| Not documented | 220 [14.81] | 214 [15.02] | -0.2 [-2.26, 2.68] |
| HR status at mBC diagnosis, n [%] | | | |
| Positive | 1117 [75.22] | 1073 [75.30] | -0.08 [-3.28, 3.16] |
| Negative | 310 [20.88] | 296 [20.77] | 0.1 [-3.08, 3.08] |
| Results unknown/not assessed | 58 [3.91] | 56 [3.93] | -0.02 [-1.36, 1.53] |
| HER2 status at mBC diagnosis, n [%] | | | |
| Positive | 239 [16.09] | 230 [16.14] | -0.05 [-2.56, 2.57] |
| Negative | 1066 [71.78] | 1022 [71.72] | 0.07 [-3.59, 3.63] |
| Equivocal | 90 [6.06] | 86 [6.04] | 0.03 [-1.65, 1.61] |
| Results unknown/not assessed | 90 [6.06] | 87 [6.11] | -0.04 [-1.68, 1.84] |
| BRCA status (germline), n [%] | | | |
| Positive | 45 [3.03] | 44 [3.09] | -0.06 [-1.11, 1.25] |
| Negative | 270 [18.18] | 260 [18.25] | -0.06 [-2.79, 2.9] |
| Other result | 9 [0.61] | 9 [0.63] | -0.03 [-0.53, 0.57] |
| Results unknown/not assessed | 1161 [78.18] | 1112 [78.04] | 0.15 [-3.4, 2.69] |
| Median number of visits [IQR] | 37 [15, 72] | 37 [15, 72] | 0 [-2, 4] |
| Median number of lines of therapy [IQR] | 1 [1, 3] | 2 [1, 3] | -1 [-1, 1] |
| Median follow-up time from mBC diagnosis, months [IQR] | 16.23 [6.07, 34.07] | 16.77 [6.07, 34.07] | -0.53 [-1.07, 2.07] |
| Year of mBC diagnosis, n [%] | | | |
| 2011 | 167 [11.25] | 165 [11.58] | -0.33 [-1.88, 2.74] |
| 2012 | 204 [13.74] | 199 [13.96] | -0.23 [-2.51, 2.77] |
| 2013 | 209 [14.07] | 204 [14.32] | -0.24 [-2.2, 2.66] |
| 2014 | 255 [17.17] | 251 [17.61] | -0.44 [-2.32, 3.32] |
| 2015 | 263 [17.71] | 252 [17.68] | 0.03 [-2.94, 2.64] |
| 2016 | 233 [15.69] | 216 [15.16] | 0.53 [-3.09, 2.2] |
| 2017 | 149 [10.03] | 133 [9.33] | 0.7 [-2.85, 1.32] |
| 2018 | 5 [0.34] | 5 [0.35] | -0.01 [-0.4, 0.43] |

Effect on example research questions

In the first analysis, we assessed OS from the date of mBC diagnosis in the two cohorts (Figure 3). Among the 1485 patients in the Reference Standard there were a total of 751 patients with recorded all-cause death events compared to 731 such patients recorded among the 1425 patients in the MACS Cohort. The estimated median OS was 2.71 years (95% CI: 2.51, 2.93) in the Reference Standard and 2.72 years (95% CI: 2.52, 2.96) in the MACS cohort (difference in median OS: -0.01 years; 95% CI: -0.31, 0.30).

For the second analysis, we compared OS as a function of HR and HER2 status (triple negative vs. HR+/HER2-) to test if there is bias in the MACS Cohort that becomes apparent when addressing research questions in smaller cohorts (Figure 4). In the Reference Standard, the estimated median OS was 3.03 years (95% CI: 2.76, 3.27) for HR+/HER2- patients compared to 1.16 years (95% CI: 0.86, 1.39) for patients with triple negative breast cancer. In line with clinical expectations, the difference in median OS between the two subpopulations was statistically significant (log-rank p-value < 0.001). This difference in OS was replicated in the MACS Cohort (HR+/HER2-: 3.08 years, 95% CI: 2.78, 3.28; triple negative: 1.17 years, 95% CI: 0.86, 1.42; log-rank p-value < 0.001). The direct comparison of median OS for the two subpopulations between the Reference Standard and the MACS Cohort showed that the differences between the two cohorts were again clinically not meaningful. The difference in median OS among patients with HR+/HER2- mBC was -0.04 years (95% CI: -0.31, 0.30) and among patients with triple negative mBC was -0.02 years (95% CI: -0.39, 0.36).

## Discussion

EHRs are a promising source of RWD for health outcomes research, but selecting targeted cohorts of oncology patients often requires the use of information that is captured only in unstructured parts of the record. In cases where research draws on these unstructured sources, human review of each chart has traditionally been needed, but this makes studies labor intensive and slow to perform on large populations.

To address the need for more efficient cohort selection for building deidentified EHR-based research datasets, we developed MACS, a technique that uses ML to filter out patients who are unlikely to be part of a cohort. Here, we applied MACS to the example of mBC and showed that it can reduce the number of charts needing human review by 78% while missing only 4% of cohort-eligible patients.

In this work, we have advanced the methods previously developed for automating cohort selection by supplementing traditional performance measures such as sensitivity and AUC with a thorough examination for bias. To perform this bias analysis, we compared the MACS Cohort to a reference standard cohort that was built using human review of each chart. First, we compared baseline clinical and demographic characteristics and found no large differences. Second, we compared the analyses of example research questions and showed that the findings did not depend on which cohort was used. This bias analysis step instilled confidence that the ML model did not systematically miss patients in a way that introduced clinically meaningful bias.

Notably, although MACS with Bias Analysis can automate a significant amount of cost and effort, it still requires the manual abstraction of unstructured data. For MACS, this abstraction is needed to produce the ground-truth labels for training and testing; for bias analysis, it is needed to extract many of the patient-level variables included in the baseline comparisons of cohorts and example analyses. Thus, even though MACS with Bias Analysis can rapidly scale cohort selection, it is built on a foundation of human expertise and on the technology to leverage that expertise at scale. Over time, we expect that automated methods will continue to reduce the manual work required for expert data curation, but this must be done carefully to ensure that quality is maintained when the downstream datasets are used for research.

The amount of bias that is permissible depends on how the data are being used, requiring some clinical and statistical judgement. When bias is deemed too great, there are at least two options that can be pursued. First, the score threshold can be adjusted to increase the sensitivity of the model. While this will not change the bias inherent in the algorithm, it means that fewer patients will be excluded, and therefore the magnitude of the bias will likely be reduced. Second, the training algorithm can be modified to address the bias directly (as long as when bias is reassessed, it is done on fresh test data). For example, higher weights can be given to patients belonging to a subpopulation that the bias analysis indicates would be systematically excluded otherwise.

Abstractors trained to do manual curation are a constrained resource. As shown here, MACS can significantly reduce the time and cost to create deidentified research datasets based on EHR data. With the savings from techniques like MACS, we can reallocate resources to other high-value abstraction tasks and to scaling research datasets in other disease areas. Thus, the ultimate benefit of MACS is that it allows for an increase in the amount and scope of research that can be done with EHR data.

**Limitations**

A limitation of our work is that our methods for bias analysis do not fully preclude the possibility of bias entering into the cohort. First, we can assess bias only for variables that are captured in the dataset. Second, even for those variables, bias may be apparent only when the joint distribution of a subset of variables is examined, and we cannot examine every possible subset. Similarly, we cannot test every example research question. Finally, the use of the reference standard cohort to assess bias may mask bias that already exists in the reference standard itself. Ultimately, our method for bias analysis relies on clinical judgement to choose variables and example analyses that will sufficiently raise confidence in the representativeness of the resulting data set.

Another limitation is that our method for generating a reference standard is inefficient when the target population has a low prevalence since we rely on randomly sampling patients of potential interest. Many patients would need to be abstracted to create a reference standard of sufficient size. In our example of mBC, the prevalence was 9%, which meant finding the 1485 patients for our reference standard required abstracting charts for over 15,000 patients. For a target population with a substantially lower prevalence, this approach to building a reference standard may become infeasible.

A final limitation is that our implementation of MACS for mBC had a somewhat decreased sensitivity for patients who were diagnosed with breast cancer at an early stage or who became metastatic only recently (Table 3). A reason for this may be that our approach used a linear model that does not include any representation of the longitudinality of the input EHR. This limitation could be addressed by exploring a non-linear model or by modeling time more directly. We left this for future work because overall, the differences in baseline distributions were small.

**Future work**

The use of MACS outlined here was for one-time cohort selection, but MACS can also support EHR-based registries when applied longitudinally. Such applications face the challenge of evolving data inputs. For example, the contents of the EHR for an individual patient will grow as that patient continues to receive care, and the population-level distribution of variables may shift as the underlying database grows or shrinks, documentation patterns change, and standards of care evolve. In such situations, both the training and test sets must be grown continuously to remain representative, and new models must be periodically deployed.

MACS can also be used to identify research cohorts beyond mBC. Within oncology, it can be applied to identify patients who have particular biomarkers, who have received a specific therapy, or who have advanced or metastatic disease in other tumor types. Beyond oncology, similar approaches can be applied to generate research cohorts in a wide variety of disease states. In general, MACS can improve the efficiency of cohort selection in situations where unstructured data is required and when patients who satisfy the cohort inclusion and exclusion criteria are relatively rare, as long as sufficient labeled data is available and the model's performance and potential for introducing bias are monitored.

## Conclusion

This work shows how combining automated ML approaches with technology-enabled abstraction and rigorous validation analyses offers a powerful method for overcoming some of the challenges of using large-scale EHR data for research. Applying MACS with Bias Analysis, it is possible to perform cohort selection for EHRs more efficiently while still preserving confidence in data quality, allowing for the acceleration of outcomes research using EHR data.

## ACKNOWLEDGMENTS

We thank Carrie Bennette and Peter Gabriel for feedback on this research and Caroline Nightingale for project management support. We also thank Nicole Lipitz and Julia Saiz-Shimosato for editing support. We also thank Amy P. Abernethy, MD, PhD, for her contributions while Chief Medical Officer and Chief Scientific Officer of Flatiron Health.

## FUNDING AND COMPETING INTERESTS

This study was sponsored by Flatiron Health Inc., which is an independent subsidiary of the Roche group. For the study period, all authors report employment at Flatiron Health, equity ownership in Flatiron Health, and stock ownership in Roche.

## AUTHOR CONTRIBUTIONS

Study design: B.B., N.N., K.S.R., M.A., M.E. Data collection: B.B., K.S.R., M.E., E.E., J.H., L.H., P.L., P.R. Data analysis: B.B., K.S.R., M.E. Data interpretation: B.B., N.N., K.S.R. Writing and review of manuscript: All

**FIGURES**

Figure 1. Schema of MACS with Bias Analysis. A. Two-step MACS cohort-selection process: in the first step, a model predicts which patients are eligible for cohort inclusion. In the second step, human abstractors manually assess patients deemed cohort-eligible by the model for cohort eligibility. B. Bias Analysis compares the cohort produced by MACS to a reference cohort in which every inclusion or exclusion decision was made by a human abstractor.

Figure 2. ROC curve of the logistic regression model used in MACS (AUC: 0.976, 95% CI: 0.973, 0.979).

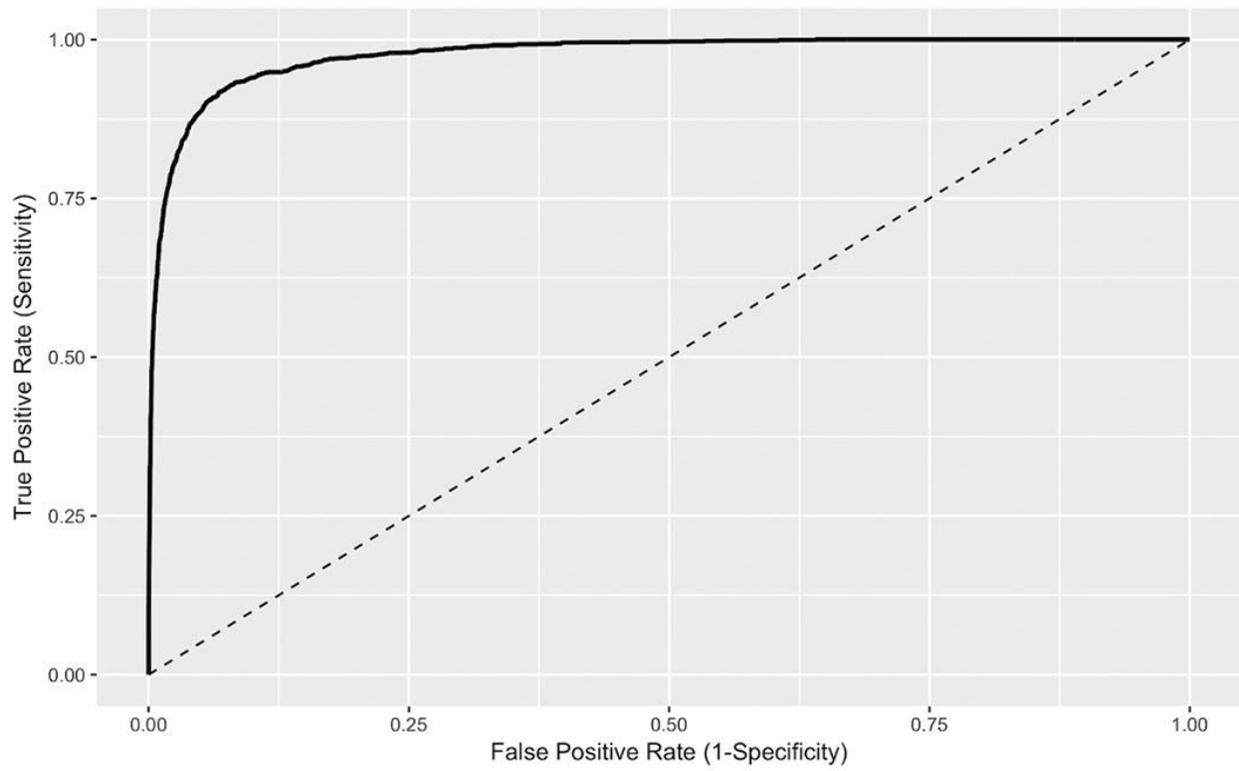

Figure 3. Kaplan-Meier estimates of OS from the date of mBC diagnosis. A. Reference Standard. B. MACS Cohort.

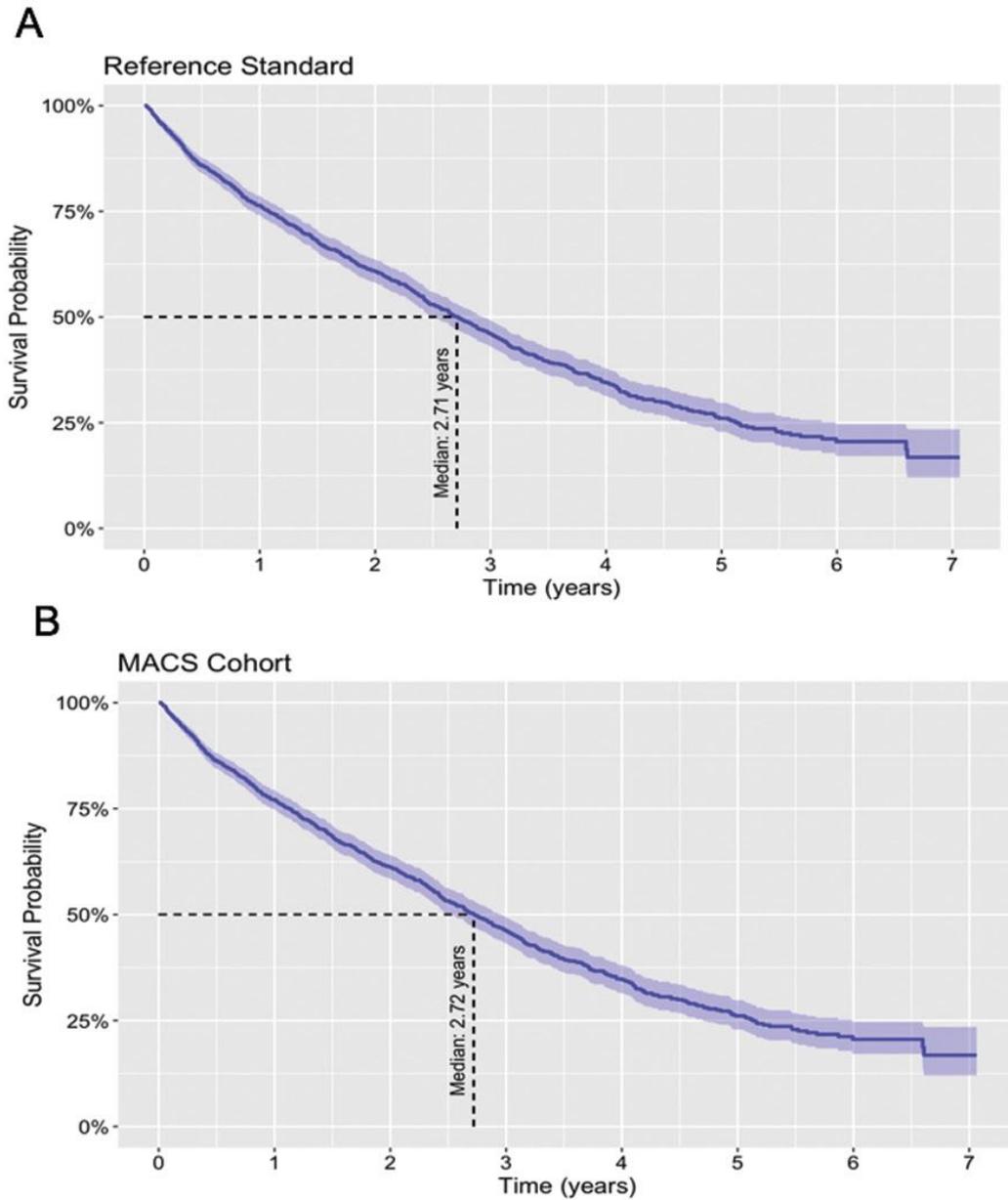

Figure 4. Kaplan-Meier estimates of OS as a function of HR and HER2 status (triple negative vs. HR+/HER2- subgroups). A. Reference Standard. B. MACS Cohort.

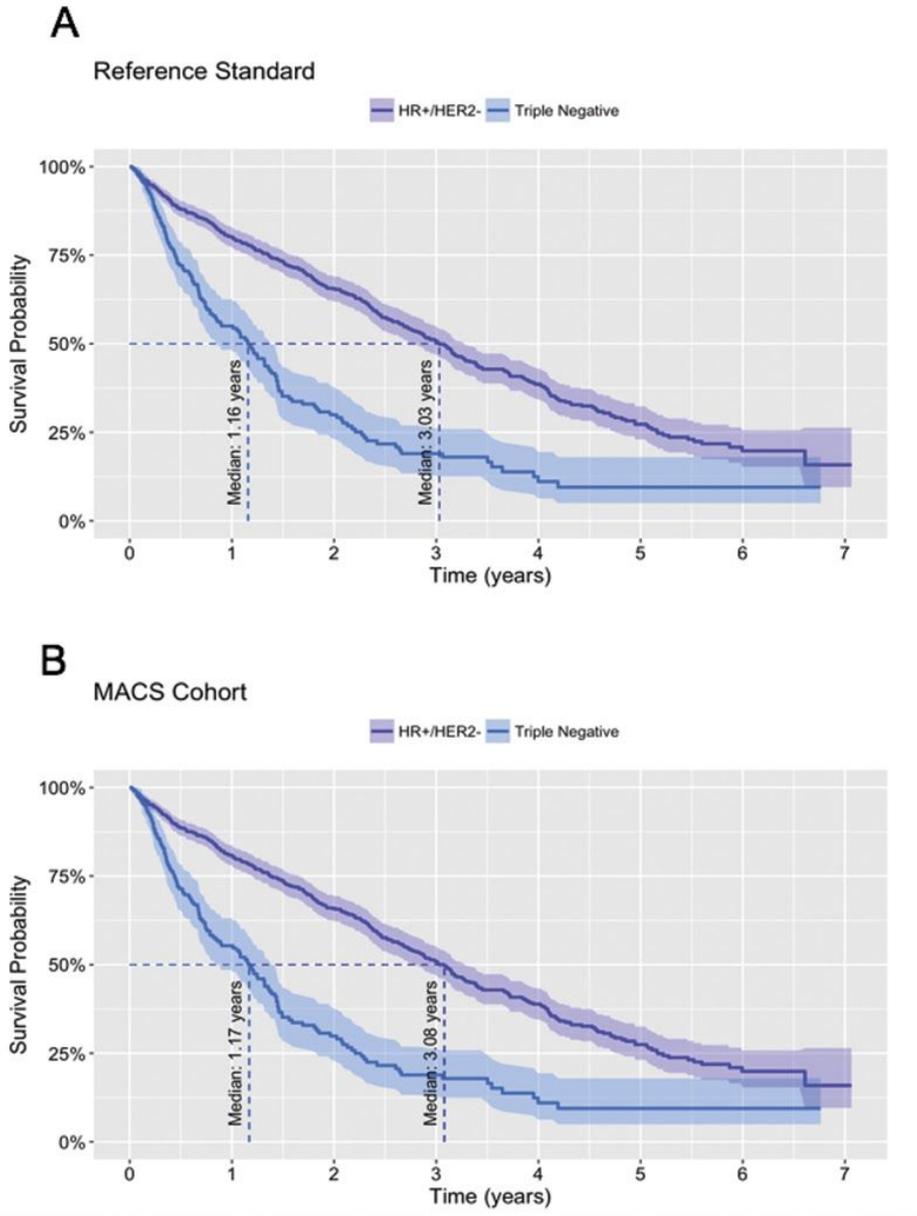